\documentclass[]{aa} % for a referee version
\usepackage{graphicx}
\usepackage{txfonts}
\usepackage[]{natbib}
\begin{document}

\title{Water and Methanol Ice in L\,1544 \thanks{Based on data
    collected by SpeX at the Infrared Telescope Facility, which
    is operated by the University of Hawaii under contract
    NNH14CK55B with the National Aeronautics and Space
    Administration.}}

\author{Miwa Goto\inst{1},
       A. I. Vasyunin\inst{2, 3},
       B. M. Giuliano\inst{4},
       I. Jim\'enez-Serra\inst{5},
       P. Caselli\inst{4},
       C. G. Rom\'an-Z\'u\~niga\inst{6},
       J. Alves\inst{7}}

\institute{Universit\"ats-Sternwarte M\"unchen,
           Ludwig-Maximilians-Universit\"at, Scheinerstr.~1, 81679
           M\"unchen, Germany %\\ \email{mgoto@usml.lmu.de}
  \and
          Ural Federal University,
          620002, 19 Mira street, Yekaterinburg, Russia \\ \email{anton@urfu.ru}
  \and
		  Visiting Leading Researcher, Ventspils International Radio Astronomy Centre,
          Inženieru 101, LV-3601, Ventspils, Latvia
  \and
          Max-Planck-Institut f\"ur extraterrestrische
          Physik, Giessenbachstrasse 1, 85748, Garching, Germany
  \and
          Centro de Astrobiolog\'{\i}a,
          Instituto Nacional de T\'ecnica Aeroespacial
          Ctra de Torrej\'on a Ajalvir, km 4
          28850 Torrej\'on de Ardoz, Madrid, Spain
   \and
           Universidad Nacional, Auton\'oma de M\'exico
           Km 107 Carretera Tijuana-Ensenada
          22870 Ensenada, BC, Mexico
   \and
          Universit\"at Wien, Department of Astrophysics
          T\"urkenschanzstra{\ss}e 17 (Sternwarte), 1180 Wien, Austria
}
\date{\today}
  % 5 {} token are mandatory

\abstract
  % context heading (optional)
% {} leave it empty if necessary
   {Methanol and complex organic molecules have been found in
     cold starless cores, where a standard warm-up scenario
     would not work because of the absence of heat sources.  A
     recent chemical model attributed the presence of methanol
     and large organics to the efficient chemical desorption
     and a class of neutral-neutral reactions that proceed fast at low
     temperatures in the gas phase.}
   % aims heading (mandatory)
   {The model calls for a high abundance of methanol ice at
     the edge of the CO freeze-out zone in cold cloud cores.}
  % methods heading (mandatory)
   {We performed medium resolution spectroscopy toward 3 field
     stars behind the starless core L\,1544 at 3\,$\mu$m to
     constrain the methanol ice abundance and compare it with
     the model predictions.}
   % results heading (mandatory)
   {One of the field stars shows a
   methanol-ice abundance of 11\,\% with respect to water
   ice. This is higher than the typical methanol abundance previously found in cold
   cloud cores (4\,\%), but is 4.5 times smaller than
   predicted. The reason for the disagreement between the observations
   and the model calculations is not yet understood.}
  % conclusions heading (optional), leave it empty if necessary
    {}

%===================================================================
% http://www.aanda.org/index2.php?option=com_content&task=view&id=170&Itemid=256
%           --- circumstellar matter
%           --- infrared: general
%           --- ISM: lines and bands
%           --- stars: pre-main sequence
%           --- techniques: spectroscopic }

  \titlerunning{Water and Methanol Ice in L\,1544}
  \authorrunning{Goto et al.}

   \maketitle

%-------------------------------------------------------------------
\section{Introduction}
%--------------------------------------------------------------------

Methanol is the most abundant alcohol in the universe.  It is
commonly accepted that the molecule is the end product of
successive hydrogenation of CO on dust grains
\citep{Tielens:1997IAUS..178...45T,Watanabe:2004ApJ...616..638W}.
In the vicinity of young protostars, icy mantles of the grains
are warmed up, and release the molecules back into the gas phase
\citep[e.g.][]{Garrod:2006A&A...457..927G}. Methanol is excited
in the warm envelopes and outflows around protostars
\citep{Flower:2010MNRAS.409...29F}, and creates a dense forest
of emission lines that almost covers the whole continuum
emission at sub-mm/mm wavelengths
\citep[e.g.][]{Bergin:2010A&A...521L..20B}.

The challenge to the current understanding of the interstellar
chemistry is the presence of methanol and other complex organic
molecules in cold clouds and starless cores found in sub-mm/mm
observations
\citep{Cernicharo:2012.759.,Bacmann:2012A&A...541L..12B,Jimenez-Serra:2016ApJ...830L...6J,Taquet:2017.607.,Punanova:2018.855.}. A
starless core is a compact, dense region of a molecular cloud,
with a typical radius of about 0.1~pc \citep[see][for
  review]{Bergin:2007ARA&A..45..339B}. Visual extinctions at the
central regions range from $A_V\sim$20\,mag for less developed
cores to $>$100\,mag for the densest ones. With no heating sources near
by or embedded within them, the gas and dust in these cores are cold, with typical temperatures $\sim$10\,K or lower
\citep{Crapsi:2007A&A...470..221C,Harju:2008.482..535H,
  Hocuk:2017A&A...604A..58H}. As a result, CO and other heavy molecules
containing oxygen and carbon are frozen onto dust grains toward
the center of these cores, possibly leaving only molecular hydrogen
and nitrogen-bearing volatile species such as NH$_3$,
N$_2$H$^+$, and HCN in the gas phase
\citep{Walmsley:2004.418.1035}.

Methanol and other large organics have been found in
starless cores near the edges of the CO depletion zones \citep{
  Tafalla:2006A&A...455..577T,Bizzocchi:2014A&A...569A..27B,
  Vastel:2014ApJ...795L...2V,Jimenez-Serra:2016ApJ...830L...6J,Chacon-Tanarro:2018arXiv180809871C}.
This is inexplicable for two reasons. First, as no energetic
sources are in cold starless cores, it is hard to eject
methanol and other molecules formed on the grain surface to the
gas phase. Second, radical-radical reactions on the grain
surface are supposed to play a key role in the formation of
complex organic molecules in the interstellar medium
\citep{Garrod:2006A&A...457..927G,Garrod:2008ApJ...682..283G}.
For the surface reactions between radicals, the dust temperatures should be 30-40\,K \citep[e.g.][]{Tielens:1982A&A...114..245T}, so that the grains are not too hot to evaporate the molecules, but also not too cold as to hinder the surface diffusion and to allow surface atomic hydrogen to saturate radicals. Although methanol
itself can form since atomic hydrogen remains mobile at low
temperatures, most of the precursors of complex organic
molecules would be fixed on their binding sites at temperatures
near 10\,K.

%==============================================================
%  Figure 1 map
%-------------------------------------------------------------
    \begin{figure*}
   \centering \includegraphics[width=1.0\textwidth]{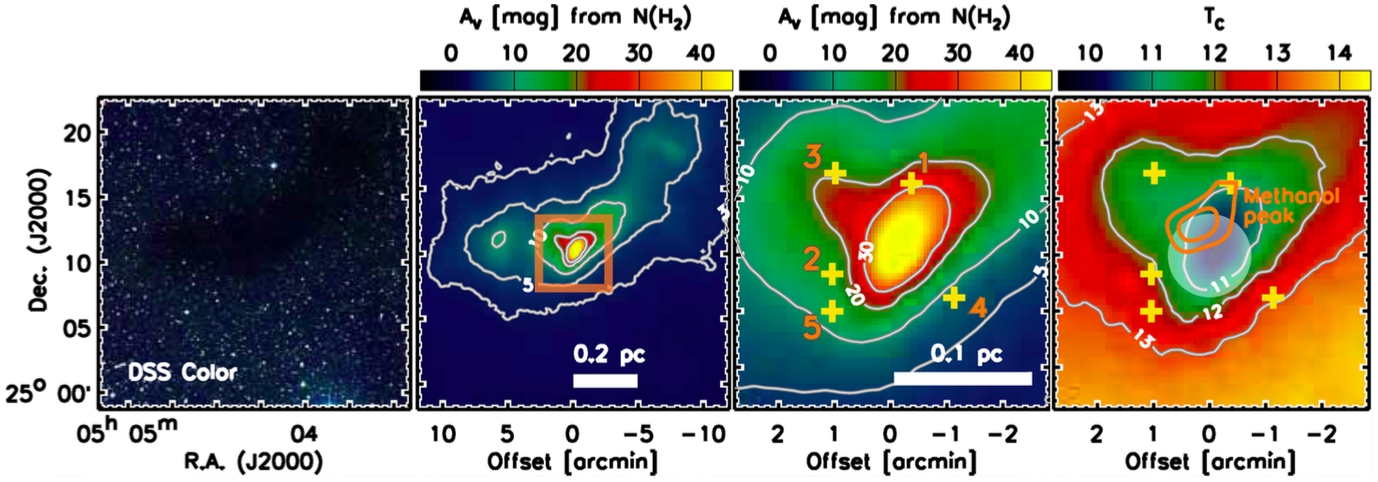}
   \caption{Background stars observed toward L\,1544 in
     the present study. First panel from the left: DSS color
     image of L\,1544 to delineate the dark patch of the
     starless core. Second: the extinction map based on the
     Herschel/SPIRE far-infrared imaging. Third: a close-up view
     of the central part of L\,1544, enclosed within the orange
     rectangle on the second panel. The positions of the
     2MASS/WISE sources are marked with crosses. Fourth: dust
     color temperature based on Herschel/SPIRE imaging. The
     methanol peak observed by
     \citet{Bizzocchi:2014A&A...569A..27B} is shown with
     orange contours. The half-transparent circle roughly
     denotes the region where the CO is frozen out
     \citep{Caselli:1999ApJ...523L.165C}.
\label{map}}
   \end{figure*}
%==============================================================

%==============================================================
%  Table 1
%-------------------------------------------------------------
\begin{table*}
\scriptsize
\caption{Column densitices of water and methanol ice.\label{t1}}
\begin{tabular}{ll ll rc c c c c c c c}
\hline \hline
\# & \multicolumn{1}{c}{WISE ID} & \multicolumn{2}{c}{(J2000)}
& \multicolumn{1}{c}{$W1$} & \multicolumn{1}{c}{$A_V$(Her)} & \multicolumn{1}{c}{$A_V$(SED)} & \multicolumn{1}{c}{$A_V$(CE)} & \multicolumn{1}{c}{Int.}
& \multicolumn{1}{c}{$\tau_{\rm 3.0}$} & \multicolumn{1}{c}{$N_{\rm H_2O}^{\rm ice}$}
& \multicolumn{1}{c}{$N_{\rm CH_3OH}^{\rm ice}$}  & \multicolumn{1}{c}{CH$_3$OH$^{\rm ice}$} \\
  &                         & \multicolumn{1}{c}{R.A.} & \multicolumn{1}{c}{Dec.}
& \multicolumn{1}{c}{[mag]} & \multicolumn{1}{c}{[mag]} & \multicolumn{1}{c}{[mag]} & \multicolumn{1}{c}{[mag]} & \multicolumn{1}{c}{[min.]}
&                           & \multicolumn{1}{c}{[10$^{18}$\,cm$^{-2}$]}
& \multicolumn{1}{c}{[10$^{17}$\,cm$^{-2}$]} & \multicolumn{1}{c}{[\%]}   \\
\hline
1 & J050415.91+251157.3 & 05:04:15.91 & +25:11:57.3 & 11.75 & 27.3 & 33.9 & 36.4 & 146 & 1.99 $^{+0.27}_{-0.23}$ & 3.33$^{+0.44}_{-0.38}$ & $<$4.0 & $<$12.1 \\
3 & J050421.72+251208.6 & 05:04:21.72 & +25:12:08.6 & 9.38 & 19.2 & 21.4 & 26.9 & 52 & 1.21 $^{+0.03}_{-0.03}$ & 2.03$^{+0.05}_{-0.05}$ & 2.2$^{+0.2}_{-0.2}$ & 10.6 $\pm$ 0.7  \\
5 & J050422.09+250937.8 & 05:04:22.09 & +25:09:37.8 & 11.60 & 12.4 & 8.9 & 15.9 & 58 & 0.67 $^{+0.17}_{-0.12}$ & 1.12$^{+0.28}_{-0.21}$ & $<$6.6 & $<$59.0  \\
\hline
\end{tabular}
\tablefoot{ID : Numbering of sources in Figure~\ref{map}.\\
$W1$  : Photometry from AllWISE catalog at the wavelength $\lambda$=3.4\,$\mu$m.\\
$A_V$ (Her): Visual extinction measured based on the extinction map calculated from Herschel/SPIRE images.\\
$A_V$ (SED): Visual extinction measured based on the comparison of the infrared SEDs of the objects with stellar photospheric model.\\
$A_V$ (CE): Visual extinction measured based on the color-dependent extinction of the background stars in the infrared.\\
Int. : On-source integration time in minutes.\\
$\tau_{\rm 3.0}$ : Peak optical depth of water ice at 3.0\,$\mu$m. The uncertainties are for 1\,$\sigma$.\\
$N^{\rm ice}_{\rm H_2O}$ : Column density of water ice. \\
CH$_3$OH$^{\rm ice}$ : Fraction of methanol ice with respect to water ice. The upper limits are for 3\,$\sigma$.\\
$N^{\rm ice}_{\rm CH_3OH}$ : Column density of methanol ice.}
\end{table*}

%==============================================================

\citet{Vasyunin:2017ApJ...842...33V} further develops the scenario proposed in~\citet[][]{VasyuninHerbst13}, and explains the presence of
methanol and of large organics in starless cores with two new
mechanisms: efficient chemical (=reactive) desorption from
CO-rich ice surface \citep{Minissale:2016A&A...585A.146M} as
opposed to water-ice surface, and a class of OH radical
reactions that proceed fast at low temperatures in the gas phase
\citep{Shannon:2013NatCh...5..745S,Balucani:2015.449.,Antinolo:2016ApJ...823...25A}.
The formation of methanol and complex organic molecules proceeds
as follows: (i) CO starts to freeze out catastrophically at
distances of thousands AU from the cloud center
\citep[e.g.][]{Caselli:1999ApJ...523L.165C}. The grain surfaces
are covered by CO-rich ice, where CO molecules are efficiently
transformed into CH$_3$OH via hydrogenation reactions. (ii) A fraction of
the CH$_3$OH molecules formed on CO-rich ice desorbs from the
surface using the extra energy made available upon formation
\citep[e.g.][]{Garrod:2006FaDi..133...51G,Minissale:2016A&A...585A.146M}.
% and to the fact that CO- and CH$_3$OH-rich surfaces makes the
% reactive desorption more efficient
% \citep{Minissale:2016A&A...585A.146M}.
(iii) Once in the gas phase, methanol and other radicals such as
CH$_3$O and HCO initiate a series of reactions in the gas phase that lead to
oxygen-bearing organic molecules. Neutral-neutral reactions
involving OH are particularly important, because they proceed
fast at low temperatures, as the lifetime of the hydrogen-bonded
intermediate complexes become longer, and the quantum tunneling
allows reactions to proceed.
% \citet{Vasyunin:2017ApJ...842...33V}
The new model not only explains how methanol and complex organic
molecules could be present in cold starless cores, but also why
they are found near the edges of the CO depletion zones \citep{Tafalla:2006A&A...455..577T,Bizzocchi:2014A&A...569A..27B}. Note that models invoking non-diffuse grain surface chemical processes have recently been proposed \citep{Jin_2020}. However, the abundances of complex organics predicted by these models fall short by at least one order of magnitude with respect to those measured toward the L1544 starless core.

The chemical model of \citet{Vasyunin:2017ApJ...842...33V} is
based on the radial profiles of density, temperatures and $A_V$
computed for the starless core L\,1544 by
\citet{Keto:2010MNRAS.402.1625K}. CO is heavily depleted in
L\,1544, showing a cavity of C$^{17}$O emission with a radius of
$\simeq$6500\,AU \citep{Caselli:1999ApJ...523L.165C}.
\citet{Bizzocchi:2014A&A...569A..27B} found that methanol
emission peaks about 4000\,AU away from the dust emission peak,
close to the rim of the CO freeze-out zone \citep[see
  also][]{Punanova:2018.855.}. Other organics such as dimethyl
ether and methyl formate were also found near the methanol peak
\citep{Jimenez-Serra:2016ApJ...830L...6J}.
\citet{Vasyunin:2017ApJ...842...33V} predicted that the local
fractional abundance of methanol ice to water ice could be as
high as 50\,\% at the edge of the CO depletion zone.  The goal
of the present study is to measure the abundances of methanol and
water ice toward L\,1544 via 3\,$\mu$m spectroscopy, and compare
them to the model predictions by \citet{Vasyunin:2017ApJ...842...33V}.
The paper is organised as follows. The observations are described in section~\ref{observation}, the
results are presented in section~\ref{result}, and the comparison between
the observations and the model is discussed in
section~\ref{discussion}.
%================================================================

%--------------------------------------------------------------
\section{Observations\label{observation}}
\subsection{Target Selection}

The gas phase methanol peak in L1544 is located about half an arcminute northeast
of the sub-mm/mm dust emission peak \citep[R.A.=05:04:17.21,
  Dec.=$+$25:10:42.8;
  J2000][]{Ward-Thompson:1994MNRAS.268..276W,Ward-Thompson:1999MNRAS.305..143W}.
In order to perform the spectroscopy through the cloud, we need
suitable background stars. We selected all stars within
5\arcmin~of the continuum peak from the WISE\footnote{The
  Wide-field Infrared Survey
  Explore. \citet{Wright:2010AJ....140.1868W}.} and
2MASS\footnote{The Two Micron All Sky Survey.
  \citet{Skrutskie:2006AJ....131.1163S}.} catalogues.  The
archival images by Herschel/SPIRE are used to construct the
extinction map on L\,1544 to compare with the locations of the
potential targets.  Five sources are visible in the 7 near-mid
infrared bands from 1.2 to 22\,$\mu$m (Figure~\ref{map}). To
avoid the sources on the foreground, the infrared photometry of
the sources are compared to the stellar photospheric models of
\citet{Bressan:2012MNRAS.427..127B} after applying the
interstellar extinction law derived by
\citet{Boogert:2011ApJ...729...92B}. All five sources are
consistent with being $G$ to $M$-type stars on the background.
The three brightest sources in WISE $W1$ band (3.4\,$\mu$m), 3,
5, and 1 in Figure~\ref{map}, are observed in the present
study. The summary of the targets is shown in Table~\ref{t1}.

% measured by the SED analysis are consistent with the dust
% column density calculated from the Herschel/SPIRE archival
% observation on L\,1544 as well.

% The visual extinctions required to red the photospheric models
%  to match the observed SED are consistent with the dust column
% density calculated from

%==============================================================
%  Figure 2 ice spectra of all
%-------------------------------------------------------------
   \begin{figure}
   \centering \includegraphics[width=0.45\textwidth]{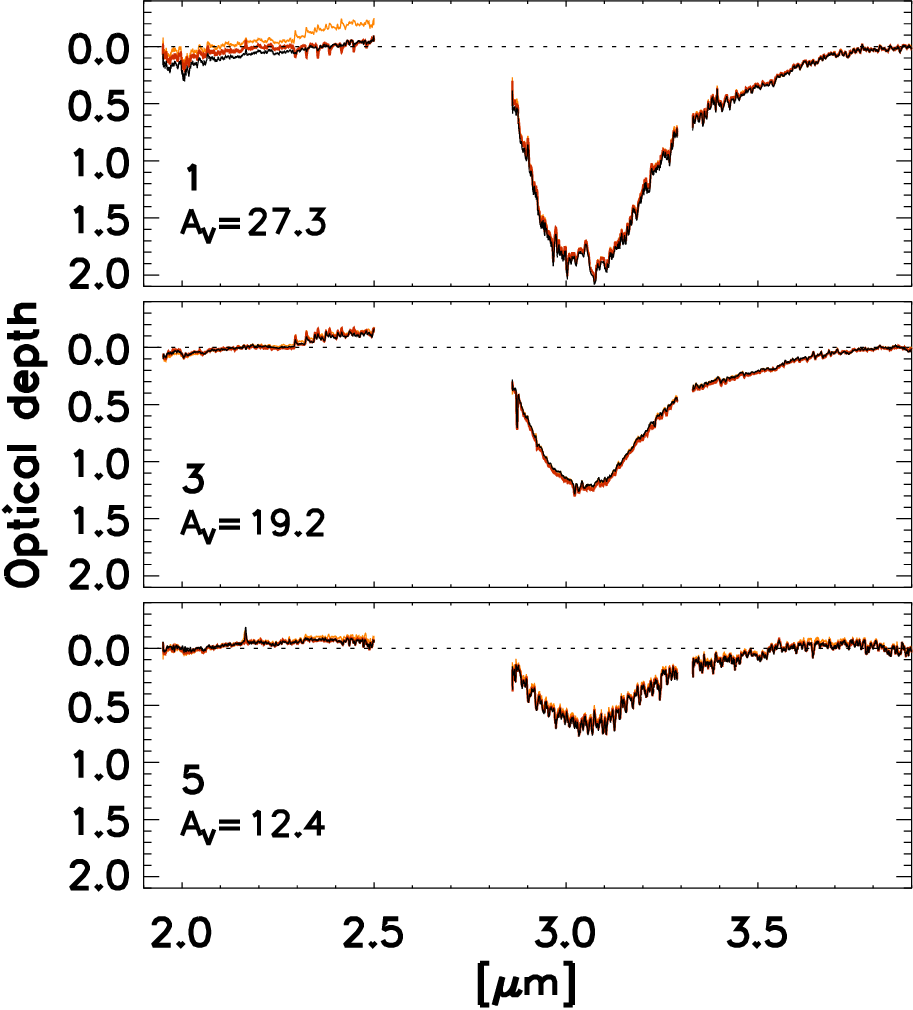}
   \caption{The optical depth spectra of 3 sources behind
     L\,1544 shown in the order of visual extinction. The
     optical depth is measured with respect to the matching
     template spectra chosen from the IRTF spectral Library
     \citep{Rayner:2009ApJS..185..289R}.  The spectra reduced
     with three best matching templates are shown (black,
     orange, and yellow traces in the order of goodness of
     matching). The identification of the sources
     is the same as in Figure~\ref{map}. The atmospheric
     transmission is low at the wavelengths 2.50--2.85\,$\mu$m
     and 3.29--3.31\,$\mu$m because of the telluric water and
     methane absorption, respectively. The data in these
     intervals are removed from the presentation. \label{ice}}
   \end{figure}
%==============================================================
%--------------------------------------------------------------
\subsection{Spex at IRTF}

The spectroscopic observations were carried out with the SpeX
spectrograph \citep{Rayner:2003PASP..115..362R} at the IRTF on
Maunakea on the nights of 20 and 21 November 2018 UT. The
instrument was remotely operated from Munich in Germany. The sky
was clear on both nights with the seeing 0\farcs5--0\farcs6 at
$K$ band. The secondary mirror of the IRTF was fresh from
recoating which helped to suppress the telescope emissivity
during the observation.

The instrument optics was set to {\tt LXD short} configuration
with 0\farcs5 $\times$ 15\arcsec~slit to deliver the spectral
resolution $R$=1200. The full coverage of the wavelength in {\tt
  LXD short} is 1.67 to 4.22\,$\mu$m. The targets were
continuously guided in the slit during the integration on the
guiding camera of SpeX at $K$ band. The position angle of the
slit was set roughly along the parallactic angle at the time of
the observation to minimize the flux loss by the atmospheric
refraction. The telescope was nodded along the slit every other
exposures to remove the background emission of the sky. The
total on-source integration times are summarised in
Table~\ref{t1}. The spectroscopic standard star HR\,1791
(B7\,III, $R$=1.66\,mag) was observed each night at a similar
airmass with the science targets. The spectrograms of the
flat-field illumination and the arc lamp were obtained after the
science and the standard star observations without changing the
telescope pointing to prevent the wavelength mapping moving on
the detector by the instrument flexure. In addition to the
calibrations at each pointing positions, extra flat field frames
were obtained in the daytime after the observations with the
telescope in the stow position.

%==============================================================
%  Figure 3 methanol P3 close up
%-------------------------------------------------------------
   \begin{figure*}
   \centering \includegraphics[width=1.0\textwidth]{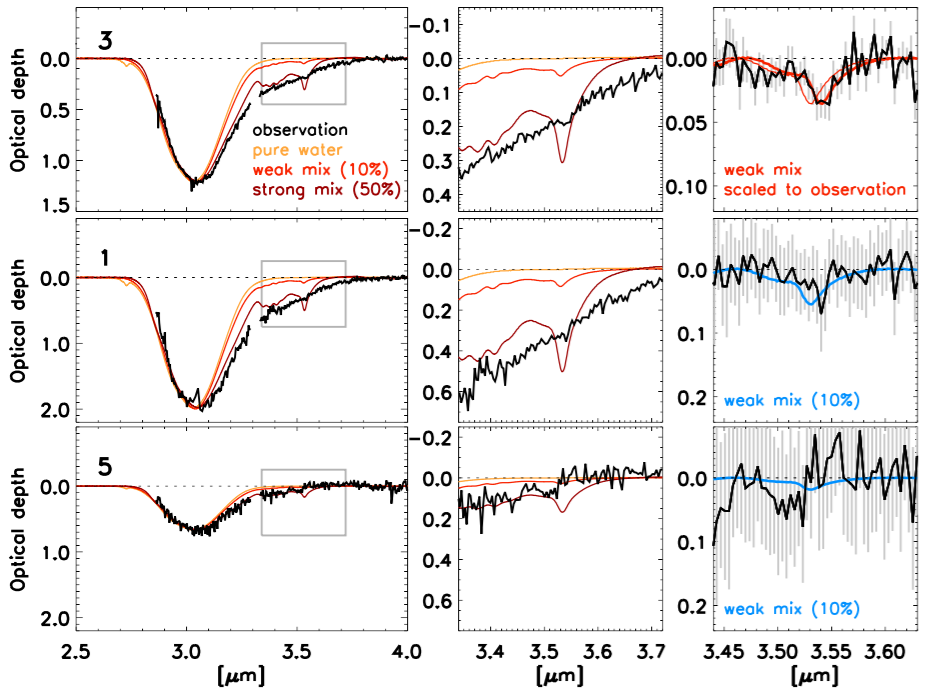}
   %\centering \includegraphics[width=1.0\textwidth]{f3b.eps}
   %\centering \includegraphics[width=1.0\textwidth]{f3c.png}
   \caption{Optical depth spectra obtained toward the 3 stars located behind
     L\,1544, shown in order of visual extinction. The
     optical depth is measured with respect to the matching
     template spectra chosen from the IRTF spectral Library
     \citep{Rayner:2009ApJS..185..289R}.  Source labels are
     the same as in Figure~\ref{map}. The
     atmospheric transmission is low at the wavelengths between
     2.50--2.85\,$\mu$m and 3.29--3.31\,$\mu$m because of the
     telluric water and methane absorption, respectively. The
     data in these intervals are removed from the presentation.
     Top row: Left panel - spectrum of star \#3 (in black
     trace) compared to the laboratory ice spectra from
     \citet{Hudgins:1993ApJS...86..713H}. The pure water ice
     spectrum is shown in yellow, the interstellar weak mix
     (H$_2$O:CH$_3$OH:CO:NH$_3$=100:10:1:1) in orange, and the
     interstellar strong mix, 100:50:1:1, in red. Middle panel -
     zoom-up view of the left panel within the gray
     rectangle. Right panel - comparison of the astronomical (black)
     and the laboratory (red) spectra after subtracting the
     local continua. The laboratory spectra of methanol ice is
     scaled so that the absorption peak matches the observation.
     The thick red trace is shifted by 0.01~$\mu$m from the
     laboratory measurement (thin trace). Middle and bottom rows:
     Same as the top row but for sources \#1 and \#5,
     respectively.  The methanol ice absorption feature is not detected.
     \label{met}}
   \end{figure*}
%==============================================================
%--------------------------------------------------------------
\subsection{Data Reduction}

The raw data were reduced by the software suite {\tt
  xSpextool}  coded by
\citet{Cushing:2004PASP..116..362C}.  {\tt xSpextool}\footnote{\tt
  http://irtfweb.ifa.hawaii.edu/\~spex/observer/}  produces
flat-field images to normalise the pixel-by-pixel fluctuation of
the detector response, and sets up the wavelength mapping on the
detector array referring to the arc-lamp images. After
calibrating the flux and the wavelength, {\tt xSpextool} removes
the sky and telescope emissions by pair subtraction, combines
the spectrograms obtained for one target, and extracts
one-dimensional spectra from the fully-calibrated spectrogram
images.

The spectra of the science targets are divided by that of the
standard star to remove the telluric lines. The blackbody
radiation spectrum of the temperature equal to the effective
temperature of the standard star is multiplied to restore the
continuum shape of the science targets. The spectral strips of
the different diffraction orders are stitched together so that
the overlapping intervals appear smooth. All these extra tasks
after the aperture extraction are handled by the programs that
come with {\tt xSpextool} suite.

The prominent \ion{H}{I} lines of the photosphere of the
standard stars (e.g. Br\,$\gamma$, Pf\,$\gamma$) are removed by
subtracting Lorentzian profiles that fit the line shape. The
\ion{H}{I} lines that are blended with the telluric lines
(e.g. Pf\,$\delta$) are removed from the science spectrum after
the telluric lines are corrected by the standard star spectrum.

The sources we observed are background field stars of late
spectral types. In order to measure the ice optical depth, the
intrinsic photospheric features have to be removed. The water
vapour absorption at 2.5--2.7\,$\mu$m is particularly
troublesome, as it lies on top of the water ice absorption at
3.0\,$\mu$m. The photospheric lines are removed by comparing the
science spectra to the template spectra of the IRTF Spectral
Library \citep{Rayner:2009ApJS..185..289R} after the templates
are reddened by the interstellar extinction law. Best matching
templates are manually sought for by inspecting how good they
reproduce the overall continuum shape and the CO overtone band
at 2.3\,$\mu$m.  The optical depths of the ice are calculated
with the template spectrum as a reference of the null
absorption. The details of the data reduction can be found in
\citet{Goto:2018.610.}. The reduced optical depth spectra are reported in Figure$\,$\ref{ice}. The spectra look flat for the wavelength range between $\sim$2 and 2.5$\,$$\mu$m as well as towards the end of the long-wavelength wing of the water absorption feature centered at 3$\,$$\mu$m, which is an indication of the goodness of the continuum fit in the observed spectra.

% The optical depth spectra reduced with three best
% matching templates are shown in Figure~\ref{ice}. The choice of
% the template would not affect the final optical-depth of ice
% much.
% In the following analysis the spectra reduced with the
% best matching template are used.

%==============================================================
%  Figure 3 methanol P1 close up (non-detection)
%-------------------------------------------------------------
%    \begin{figure*}
%    \centering \includegraphics[width=1.0\textwidth]{f4.eps}
%    \caption{
%      In the right most
%      panel the spectrum of the interstellar weak mix with the
%      methanol ice includsion 10\,\% with respect to the water
     %      ice \citep{Hudgins:1993ApJS...86..713H}.
%      \label{met1}}
%    \end{figure*}

%==============================================================
%--------------------------------------------------------------
\section{Result \label{result}}
\subsection{Methanol ice at 3.54\,$\mu$m}

The prime goal of the observation is to detect the $\nu_3$
stretching band of methanol ice at 3.53--3.54\,$\mu$m. Among 3
stars observed, only source \#3 shows significant absorption at
3.54\,$\mu$m over the noise level (top row in
Figure~\ref{met}). To measure the methanol to water ice ratio
quantitatively, the laboratory transmission spectra of water ice
with different degrees of methanol inclusions
\citep{Hudgins:1993ApJS...86..713H} are compared to source \#3
spectrum. The interstellar weak mix of
\citet{Hudgins:1993ApJS...86..713H} is a blend of ice with the
mixing ratio H$_2$O:CH$_3$OH:CO:NH$_3$=100:10:1:1. The
interstellar strong mix only differs in containing more methanol
ice following the ratio H$_2$O:CH$_3$OH:CO:NH$_3$=100:50:1:1 (see the
left panels in Figure~\ref{met}). The pure water ice does not
contain methanol or other molecules. The laboratory data
are taken for low temperature measurements at 10\,K.

The use of a H$_2$O:CH$_3$OH:CO:NH$_3$ chemical mixture is justified by the chemical models of the L1544 starless core \citep[][]{Vasyunin:2017ApJ...842...33V}. In these models, the formation of CH$_3$OH from CO ice hydrogenation takes place while O and N atoms (still present in the gas phase) also adsorb onto the grain surface. This leads to the production of H$_2$O and NH$_3$ ice on the surface at the same time as CH$_3$OH in such a way that the presence of H$_2$O and NH$_3$ ice on CO-rich mantles is expected \citep[see][submitted]{Mueller_2020}. In addition, note that the red wing of the water absorption feature at 3$\,$$\mu$m, can be contaminated by NH$_3$ hydrates \citep{Dartois_2001}.

We employ the laboratory data from \citet{Hudgins:1993ApJS...86..713H} because they are easily available from the most common databases and because they present the advantage of allowing a straightforward comparison with ice mixtures with two significant H$_2$O:CH$_3$OH ratios (100:10 and 100:50). In particular, the data from \citet{Hudgins:1993ApJS...86..713H} are the only ones available for a H$_2$O:CH$_3$OH ratio of 100:10, required to explain our observations (see below).

% The transmission spectra are slightly adjusted so that the
% absorption on the continuum at 2.5\,$\mu$m and 4.0\,$\mu$m are
% exactly null.
The transmission spectra are converted into optical depths,
and then scaled so that the peak optical depth of water ice
matches what is observed toward source \#3 at 3.0\,$\mu$m (left
panel in the top row of Figure~\ref{met}; middle panel is a
close-up view at 3.54\,$\mu$m). Polynomial functions are fitted
to the local continua near 3.54\,$\mu$m, and subtracted from the
lab and the astronomical data (right panel). An excess
absorption is found at 3.54\,$\mu$m on the source \#3 spectrum,
with a corresponding optical depth of $\tau \sim 0.03$.  This is
taken as a positive detection of methanol ice. If we linearly
scale the laboratory spectrum of interstellar weak mix to the
astronomical spectrum, the methanol over water ice ratio is
10.6$\pm$0.7\,\%. This ratio is consistent with those measured toward other starless cores such as L\,429 and L\,694 \citep[of $\sim$12-14\%; see][]{Boogert:2011ApJ...729...92B,Chu_2020}, also using the 3.54$\,$$\mu$m methanol absorption feature.

There is a small offset (0.01\,$\mu$m) in the peak absorption
wavelengths towards longer wavelength between the laboratory and the astronomical spectra
(right panel in the top row of Figure~\ref{met}). The nature of
the offset is not clear. The size of the shift is too large to
be accounted for as an uncertainty of the wavelength calibration
in the SpeX spectra.The water ice absorptions at 3.0\,$\mu$m
have a similar shift between the laboratory and the astronomical
spectra. This shift could be produced by scattering effects
due to different grain shapes and sizes \citep[see e.g.][]{Pontoppidan_2003, Thi_2006, Noble_2013}. However, given the low signal-to-noise ratio
of the methanol absorption feature, these effects are probably not noticeable. The observed shift is more likely associated with variations in the chemical composition of the ices (W.-F. Thi, private communication). This will be explored in a separate paper \citep[][submitted]{Mueller_2020}.

We point out that a comparison with spectra from the Sackler Laboratory Ice Database \citep{Fraser_2004} has also been attempted for CH$_3$OH:CO 1:1 ice mixtures and layers. The laboratory methanol feature at 3.54$\,$$\mu$m, however, does not show a significant shift when co-deposited with CO as compared to the layered deposition, at least for the ratio probed in these experiments. In any case, the comparison with the data from \citet{Hudgins:1993ApJS...86..713H} provides a better match to the CH$_3$OH ice absorption spectra observed toward star \#3 in the L1544 starless core.

The same analysis used for star \#3 is applied to sources \#1 and \#5, and shown in
Figure~\ref{met}. Source \#1 is the closest to the methanol
peak, and is expected to show the highest methanol-ice abundance
among the 3 sources observed. However, the source is faint, and
we could only set the upper limit on the methanol-ice abundance
at 12.0\,\% for 3\,$\sigma$ significance. The upper limit of the
methanol ice abundance on source \#5 is 59\,\%.

% The expected strength of the methanol ice
% absorption is compared to the observed spectrum in the right
% panel of .

%-------------------------------------------------------------
% 3.2
\subsection{Water ice at 3.0\,$\mu$m\label{water_ice}}

The water ice is positively detected in all three sources in
L\,1544. The peak optical depths are measured by fitting a
Gaussian profile to the spectra and shown in Table~\ref{t1}. To
measure the column density of the water ice quantitatively,
first the optical-depth spectrum of pure water ice from the
laboratory \citep{Hudgins:1993ApJS...86..713H} is scaled so that
the maximum optical depth $\tau_{\rm 3.0}$ is unity.
The normalized optical depth spectrum is integrated over the whole
band to the equivalent width. The column density of the water
ice in the laboratory is then calculated from the equivalent
width by multiplying the integrated band strength of
\citet{Gerakines:1996A&A...312..289G}. This way, we know that
the unit peak optical depth at 3.0\,$\mu$m amounts to a column
density of $1.68 \times 10^{18}$\,cm$^{-2}$. The column density
of the water ice is calculated by multiplying this factor to the
observed $\tau_{\rm 3.0}$ of each source.

We note that the laboratory pure water ice spectrum does not reproduce the red wing observed for the water absorption features measured at 3$\,$$\mu$m toward the three stars. This is caused by a possible mixture with NH$_3$ ice and scattering by larger grains \citep[see Section 3.1 and][]{Dartois_2001,Pontoppidan_2003,Thi_2006,Noble_2013}. The Gaussian profile, however, nicely fits the water feature between $\sim$2.9 and 3.15$\,$$\mu$m, which is sufficient to accurately determine the peak optical depths \citep[see, for instance,][]{Chu_2020}.

%==============================================================
%  Figure 4 water ice tau vs Av
%-------------------------------------------------------------
   \begin{figure}
   \centering \includegraphics[width=0.5\textwidth]{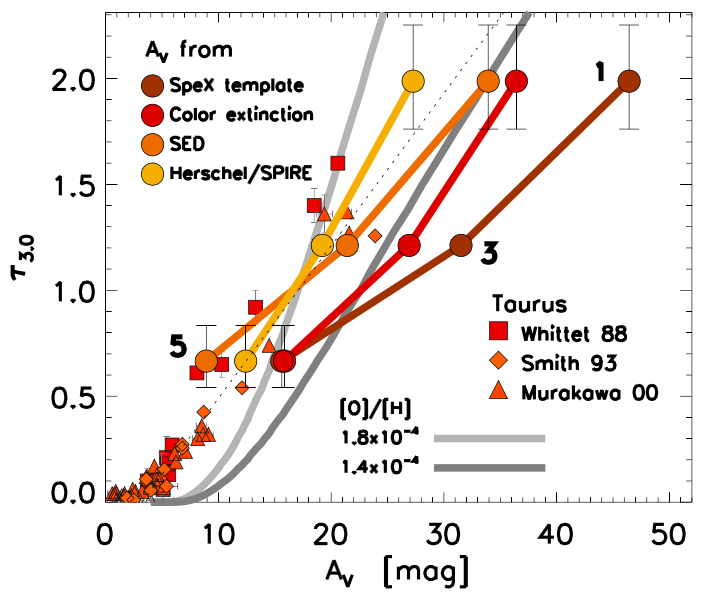}
   \caption{Peak optical depth of water ice at 3.0\,$\mu$m in
     L\,1544 plotted against the visual extinction ($A_V$)
     toward the sources. $A_V$ was measured four different ways:
     by comparing object spectrum with the IRTF spectral library
     (dark red circles), from the extinction map constructed by the colors of reddened background stars (red circles), by comparing the object SEDs with stellar photospheric model (orange circles), and from the dust
     column density calculated based on the far-infrared
     thermal emission of dust observed by Herschel / SPIRE (yellow
     circles). The optical depths of water ice on the sources in
     Taurus Molecular Cloud are shown for comparison (red
     symbols). The references are
     \citet{Whittet:1988MNRAS.233..321W,Smith:1993MNRAS.263..749S,
       Murakawa:2000ApJS..128..603M}. The dashed line denotes
     $A_V$ -- $\tau_{\rm 3.0}$ relation known in Taurus
     \citep{Whittet:2001ApJ...547..872W}. The
     range of $A_V$ for each science source measured in
     different techniques is 50--100\,\%. The peak optical
     depths of water ice in L\,1544 are in the range between
     being consistent with and smaller than those in
     Taurus. The gray lines represent the model calculation by
     \citet{Vasyunin:2017ApJ...842...33V} with standard
     low-metal abundance of oxygen ([O]/[H]=$1.8\times 10^{-4}$; light gray) and a reduced one ($1.4\times 10^{-4}$; dark gray). \label{wasser}}
   \end{figure}
%================================================================
%------------------------------------------------------------
\section{Discussion\label{discussion}}
\subsection{Assessment of $A_V$}

We discuss first how the visual extinction was measured on the
three sources observed. In the present study, $A_V$ was
estimated in four different ways.
%------------------------------------------------------------
(i) We used the thermal emission of dust grains in the far
infrared observed by SPIRE on Herschel at 250, 350 and
500\,$\mu$m. The mass absorption coefficient used was
$\kappa_\nu \propto \nu^\beta$, where $\beta=2.0$, and is
normalized to $\kappa_0 = 0.1$\,cm$^{2}$\,g$^{-1}$ at
250\,$\mu$m \citep{Hildebrand:1983QJRAS..24..267H}.
%------------------------------------------------------------
(ii) We calculated $A_V$ with the color extinction method
developed by \citet{Lombardi:2009A&A...493..735L} ({\it NICEST}
algorithm) extended for six near-infrared bands [$J$, $H$, $K$,
  IRAC\,1 (3.55\,$\mu$m), IRAC\,2 (4.49\,$\mu$m), IRAC\,3
  (5.73\,$\mu$m) and IRAC\,4 (7.87\,$\mu$m)]. The $J$, $H$ and
$K$ images were obtained with the OMEGA\,2000 camera at the
3.5\,m telescope in Calar Alto Observatory. The IRAC/Spitzer
data is from the Taurus Legacy Project\footnote{\tt
  https://irsa.ipac.caltech.edu/data/SPITZER/Taurus/\\docs/delivery\_doc2.pdf}.
The extinction map was constructed using the extinction law of
\citet{Roman-Zuniga:2007ApJ...664..357R}.
%------------------------------------------------------------
(iii) We compared the infrared SEDs of the science targets taken
from 2MASS and WISE catalogues (1.2--22\,$\mu$m)
% \citep{Skrutskie:2006AJ....131.1163S,Wright:2010AJ....140.1868W}
to the stellar photospheric model of
\citet{Bressan:2012MNRAS.427..127B} to assess the spectral types
of the stars and the extinction on the line of sights
simultaneously. The empirical extinction curve of
\citet{Boogert:2011ApJ...729...92B} was used.
%------------------------------------------------------------
(iv) We used the IRTF spectral library
\citep{Rayner:2009ApJS..185..289R} as the null absorption
reference to measure the ice optical depth. $A_V$ was measured
simultaneously in the same way with (iii). The empirical
extinction curve of \citet{Boogert:2011ApJ...729...92B} was
used.
%------------------------------------------------------------

The visual extinctions on the three sources measured by the
techniques above are listed in Table~\ref{t1}. They should all
match together on each source ideally, which is not the case
apparently. Instead of trying to reconcile the $A_V$s, we would
like to underscore the following points. (1) $A_V$ on a single
line of sight measured different ways may have a range nearly
100\,\%, i.e., if $A_V$ is 10\,mag, the full interval of the
scatter could be $\pm$5\,mag. (2) When testing solid-phase
chemistry in the interstellar medium, $A_V$ may not be a
reliable reference of the mass in front of a target. It is
preferred to have more than one species of molecules in ice
observed in a single line of sight so that $N({\rm H_2O})^{\rm
  ice}$ may serve as a reference of the solid matter, for
instance.  The complications in using $A_V$ as a reference of
mass in the line of sight in the context of ice study has been
discussed in
\citet{Chiar:2011ApJ...731....9C,Boogert:2013ApJ...777...73B,
  Whittet:2013ApJ...774..102W}.

Here we present a non-exhaustive list of the problems in
measuring $A_V$, and introduce one issue special to the present
case. A cloud may have a particular geometry, such as local
minima and maxima, or an extra extinction on foreground and
background. All are potential sources of unexpected tweak in
$A_V$ on a particular line of sight. The theoretical models of
stellar photospheres used in technique (iii) requires numbers of
inputs that are not well known for an individual target, such as
the age, the metallicity, or the presence of circumstellar
dust. The uncertainty in the photospheric reference propagates
to the uncertainty in $A_V$. The color-dependent extinction
technique (ii) requires the intrinsic color of stars as the null
reference of the extinction. The intrinsic color is often
sampled in a nearby control field. The fidelity of the intrinsic
color systematically affects the final extinction measured. The
interstellar extinction law is not universal, but depends on the
optical property of the dust grains on the line of sight, such
as grain size and chemical composition. Dust optical
properties also affect the wavelength dependency of the mass
absorption coefficient $\kappa_\nu$, and therefore the total
mass of the dust grains deduced from the far-infrared dust
emission.

Another issue is the
astrometric accuracy. The techniques (i) and (ii) do not use the
photometry or the spectroscopy of the science targets directly,
but read $A_V$ from the extinction map at the position of the
targets. The accuracy of the extinction relies on how accurate
the astrometry of the extinction map matches with the coordinate
system of the targets. The angular resolutions of the extinction
map produced by the techniques (i) and (ii) are 10\arcsec~to
1\arcmin~in most cases, while a typical starless core
($\sim$0.1\,pc) spans 2\arcmin~at 140\,pc away. Astrometric
mismatch may result in serious offsets in $A_V$ when a cloud has a
small-scale structure and a target is located close to it.

The wide range of possible $A_V$ toward our science targets makes it
difficult to compare the water-ice optical depth to the existing
observations in the Taurus Molecular Cloud
\citep[Figure~\ref{wasser};][]{Whittet:1988MNRAS.233..321W,Smith:1993MNRAS.263..749S,
  Murakawa:2000ApJS..128..603M}. The peak optical depths of
water ice $\tau_{\rm 3.0}$ in L\,1544 are either consistent with
Taurus when compared with the $A_V$ measured by SPIRE/Herschel and SED
analysis, or smaller, when compared with the $A_V$ measured using IRTF templates and the color-dependent extinction technique.

%==============================================================
%  Figure 5 Vasyunin 2017 - CH3OH/H2O ratio vs Av
%-------------------------------------------------------------
   \begin{figure}
   \centering \includegraphics[width=0.45\textwidth]{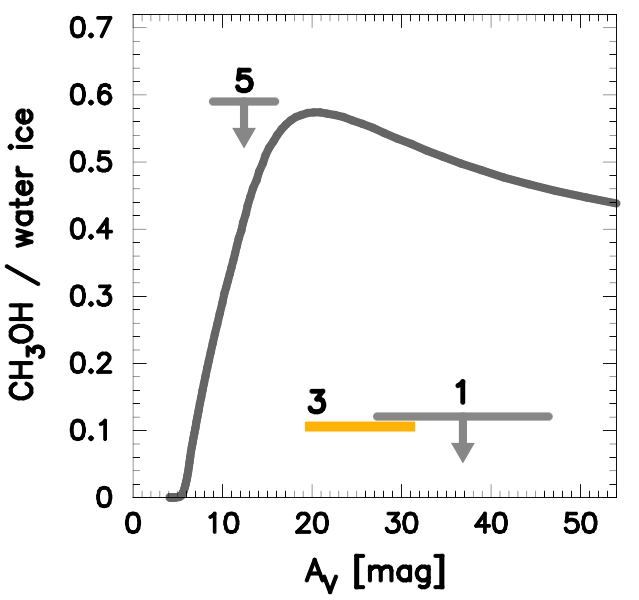}
   \caption{Fraction of methanol-ice with respect to water-ice
     column density plotted as a function of the visual
     extinction $A_V$ along the line of sight through the cloud.
     The gray curve is the model calculations by
     \citet{Vasyunin:2017ApJ...842...33V} with an updated radial density profile and a reduced initial oxygen abundance (see text for details).  The methanol to
     water ice ratio observed on source \#3 is marked by a
     horizontal yellow line, which denotes the full range of $A_V$ measured by the different
     methods. The upper limits on the methanol ice abundance are
     shown with downward arrows in light gray colours for sources \#1 and
     \#5. The fraction of the methanol ice detected toward star \#3 is
     4.5 times smaller than predicted by the models. \label{vasyu}}
   \end{figure}
%================================================================
\subsection{Comparison with the model\label{comparison}}

We will now compare the measured methanol abundance to the
chemical model calculated by
\citet{Vasyunin:2017ApJ...842...33V}. There are two
modifications from the published model of L\,1544.  First, the
gas density profile is
updated. \citet{Vasyunin:2017ApJ...842...33V} used the radial
profile of gas density calculated by
\citet{Keto:2010MNRAS.402.1625K}. This density profile turns out
to be marginally inconsistent with the far-infrared/mm emission
of dust grains \citep{ChaconTanarro:2019.623.}. The chemical
model is recalculated using the density profile proposed by
\citet{ChaconTanarro:2019.623.}. The radial profiles of the
abundances of molecules in gas and ice are affected little.

Second, the oxygen abundance is reduced. Figure~\ref{wasser}
shows the comparison of $\tau_{\rm 3.0}$-$A_V$ correlations
between the observations and the model. The curve computed with
the standard low-metal oxygen abundance ($1.8 \times 10^{-4}$; the initial
gas-phase abundance at the start of the cloud collapse) predicts
a slope that is too steep (light gray) to match the
observations. We had to reduce the oxygen abundance by 23\,\%
($1.4 \times 10^{-4}$). This is a known problem called {\it
  oxygen crisis} \citep{Ayres:2008ApJ...686..731A}. The crisis
here implies not the shortage of oxygen but that the solar
abundance of oxygen is significantly more than the total sum of the element
observationally identified in the interstellar medium
\citep{Whittet:2010ApJ...710.1009W,Jenkins:2009ApJ...700.1299J,Jenkins:2019.872.}. It is also interesting to note that \citet[][]{Hincelin_ea11} found that a similar elemental abundance of oxygen was needed in their models to reproduce the observed concentration of molecular oxygen in molecular dark clouds.
Oxygen is present in the interstellar medium in atomic and
molecular form. It is also incorporated in larger molecules,
silicates and oxides in dust grains, including ice mantles.  In
a dense cloud, observable oxygen in ice, dust and gas accounts
only 65\,\% \citep{Whittet:2007ApJ...655..332W} of the solar
abundance of oxygen [O]/[H] =$4.57 \times 10^{-4}$
\citep{Asplund:2009.47.481}. The reduced [O]/[H] abundance
decreases $\tau_{\rm 3.0}$ with respect to $A_V$, so that the model results match better the canonical slope observed in the Taurus Molecular
Cloud (dark gray trace).
% reproduce the observed optical depth of water ice, either
% (gray trace in Figure~\ref{wasser}).
The deficit of $\tau_{\rm 3.0}$ at the low $A_V$ could stem
from the chemical youth of L\,1544 compared to other
star-forming clouds in Taurus. The abundances of methanol and other large organics in the gas phase observed by \citep{Jimenez-Serra:2016ApJ...830L...6J} are still reproduced within an order of magnitude by the model with the reduced oxygen abundance, and with the peak in the abundance radial distribution matching properly the radial distance of the observed methanol peak \citet{Bizzocchi:2014A&A...569A..27B}.
The local fractional abundance of
methanol ice with respect to $n_{\rm H}$ is affected little.

The local fractional abundances of ice are converted into
column densities through the cloud in the same way as in
\citet{Jimenez-Serra:2016ApJ...830L...6J}. The model abundance
of methanol ice with respect to water ice is shown in
Figure~\ref{vasyu} as a function of $A_V$ (gray trace). The
predicted abundance of methanol ice reaches 50\,\% at $A_V =
$18\,mag, and becomes flat toward $A_V$=40-50\,mag with
30-40\,\%. Neither of the observed methanol ice fraction toward
source \#3, nor the upper limit on source \#1 come close to the
high abundances predicted, regardless of the large uncertainty
in the visual extinction on the different lines of sight.

\begin{figure}
   \centering \includegraphics[width=0.45\textwidth]{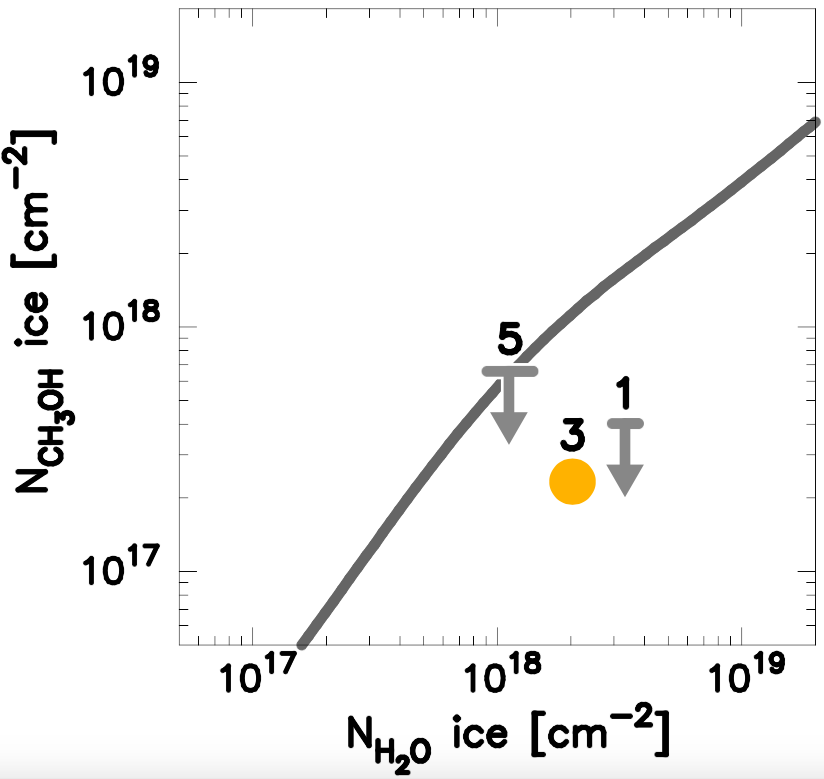}
   \caption{Column density of methanol ice plotted against that
     of water ice. The gray curve is the model calculated by
     \citet{Vasyunin:2017ApJ...842...33V} with the updated radial density profile and the reduced initial oxygen abundance (see details in the text). The methanol ice
     column density on source \#3 is shown with a yellow circle.
     The uncertainty in the column densities are smaller than
     the size of the symbol. The upper limits on sources \#1 and
     \#5 are shown by downward arrows. The extent of the
     horizontal bars denotes the uncertainty in $N({\rm
       H_2O})^{\rm ice}$. The methanol ice detected on source
     \#3 and the upper limit set on source \#1 are smaller than
     the values predicted by the model by factors of 5.0 and 4.1,
     respectively.\label{column}}
   \end{figure}

Figure~\ref{column} compares the observed ice column densities,
$N({\rm H_2O})^{\rm ice}$ and $N({\rm CH_3OH})^{\rm ice}$, to
the model. The column density of the water ice on source \#3 is
$2.03 \times 10^{18}$\,cm$^{-2}$. The column density of the
methanol ice predicted by the model is then $N({\rm CH_3OH})_{\rm ice} = 1.1\times
10^{18}$\,cm$^{-2}$, which is 5 times larger than what is
actually observed (2.2$^{+0.2}_{-0.2}$$\times$10$^{17}$$\,$cm$^{-2}$; Table~\ref{t1}). The column density of methanol in the gas phase towards L1544 is $\sim$ few $\times$~10$^{13}$~cm$^{-2}$~\citep[see e.g.][]{Bizzocchi:2014A&A...569A..27B}. Thus, the absolute majority of CH$_{3}$OH in L1544 is in the solid phase. In the model by \citet{Vasyunin:2017ApJ...842...33V}, methanol exhibits a similar behaviour.

It is not clear why the observed and the predicted
methanol ice abundances differ that much. One technical issue is
that it is not straightforward to convert the abundances of ice
given in a model to the optical-depth spectra which is the
observable.  An ice mantle is a three-dimensional structure on
the surface of a grain that works as a substrate.  The optical
constants of a single grain, in principle, should be calculated
with a realistic geometry of the mantle on the grain with the
vertical profile of the molecular abundances given in the model.
The visibility of the molecules may change depending on the
depth in the ice mantle. On the other hand, an observation assumes
that every molecule in the ice mantle contributes equally to the
ice opacity, regardless of the depths of their presence or the
size of the grains that they cover.

\citet{Vasyunin:2017ApJ...842...33V} indeed distinguishes the
surface and the bulk molecules in the ice mantle, in the sense
that molecules in the bulk, under a few layers of the ice mantle, do
not participate in the surface chemistry, but only contribute to the chemistry by slow diffusions through the ice matrix. As a consequence, the
chemical compositions of ice are substantially different in bulk
and surface. In the present study, the bulk composition is used
to compare with the observations, because it is a conservative
choice, as the methanol to water ice ratio is even higher on the
surface at the $A_V$ in question. Invisible population of
methanol underneath the thick ice mantle therefore would not
account for the deficit of the methanol ice in the present case.

We should keep in mind, however, that a chemical model is built on numerous parameters that we have not yet understood well. Rate coefficients of gas-phase reactions \citep{Wakelam:2005A&A...444..883W, Vasyunin:2004AstL...30..566V, Wakelama:2006A&A...451..551W, Vasyunin_2008}, binding energies of molecules to grain surface \citep{Wakelam:2017MolAs...6...22W}, or diffusion efficiencies on the surface \citep{Iqbal:2018...620A.109I} are known poorly, let alone the physical conditions of the
cloud in question \citep{Wakelama:2006A&A...451..551W, Wakelamb:2006A&A...459..813W,Wakelam:2010A&A...517A..21W}. As a result, the uncertainties in the output chemical abundances may exceed an order of magnitude. Finally, although it is commonly accepted that methanol in dense molecular clouds is formed efficiently only via the surface hydrogenation of CO molecules~\citep[][]{Watanabe_ea02, Fuchs_ea09}, the processes that may influence the methanol production rate are not fully understood yet~\citep[see e.g.][]{Minissale_ea16b}.
% , if the
% observation misses the water ice in the same way.

% here
% The model may have overpredicted the abundance of the methanol
% ice.
% \citet{Whittet:2001ApJ...547..872W}.
%==============================================================
%  Figure 6 Vasyunin 2017 - N(CH3OH) ice vs N(H2O) ice
%--------------------------------------------------------------
%--------------------------------------------------------------
%==============================================================

%==============================================================
\section{Conclusions}

We performed 3\,$\mu$m medium resolution spectroscopy toward 3
field stars behind the starless core L\,1544 to constrain the
methanol ice abundance with respect to the water ice. One of
them shows a methanol-ice abundance 11\,\% with respect to water.
This is consistent with the value measured toward other starless cores such as L\,429 and L\,694
\citep[of $\sim$12-14\%;][]{Boogert:2011ApJ...729...92B,Chu_2020}; however, it is still
way too low compared to the chemical model computed by
\citet{Vasyunin:2017ApJ...842...33V}. The comparison of the observations with the model reveals that the initial abundance of atomic oxygen at the start of the chemical evolution of the dense cloud in the model, has to be reduced by 23\% to match the optical depth of water ice as a function of visual extinction toward L1544.

%==============================================================
\begin{acknowledgements}

  M.G. thanks all the staff and crew of the IRTF, in particular,
  Mike Connelley, Brian Cabreira, Dave Griep, Miranda
  Hawarden-Ogata and Bobby Bus who kindly helped conducting the
  observation from Munich. We would like to thank the
  hospitality of the Hawaiian community that made the research
  presented here possible. My sincere appreciation goes to Jorma
  Harju and Viktor Zivkov who gave me thorough instructions how
  to construct dust emission / color temperature maps, and a
  color extinction map, respectively. We thank Wing-Fai Thi for the fruitful discussion about the influence of dust grain size on the shape of methanol ice absorption feature. This publication makes use
  of data products from the Wide-field Infrared Survey Explorer,
  which is a joint project of the University of California, Los
  Angeles, and the Jet Propulsion Laboratory/California
  Institute of Technology, funded by the National Aeronautics
  and Space Administration. This publication makes use of data
  products from the Two Micron All Sky Survey, which is a joint
  project of the University of Massachusetts and the Infrared
  Processing and Analysis Center/California Institute of
  Technology, funded by the National Aeronautics and Space
  Administration and the National Science Foundation. The
  optical spectra of the ice was taken from Sackler Laboratory
  Ice Database. This research has made use of NASA's
  Astrophysics Data System. This research has made use of the
  SIMBAD database, operated at CDS, Strasbourg, France. M.G. is
  supported by the German Research Foundation (DFG) grant GO
  1927/6-1. The work of A.I.V. is supported by the Russian Ministry of Science and Higher Education via the State Assignment project FEUZ-2020-0038. A.I.V. is a head of the Max Planck Partner Group at the Ural Federal University. I.J.-S. has received partial support from the Spanish FEDER (ESP2017-86582-C4-1-R) and the State Research Agency (AEI; PID2019-105552RB-C41). C.R.-Z. acknowledges support from program UNAM-DGAPA IN112620, Mexico. The authors thank the anonymous referee for the valuable suggestions that have significantly improved the manuscript.

\end{acknowledgements}
%==============================================================
\bibliographystyle{aa} % style aa.bst
\bibliography{aa}

%==============================================================
\end{document}